\documentclass[a4paper,12pt]{article}
\usepackage{graphicx}

\def\ds{\displaystyle}
\def\beq{\begin{equation}}
\def\eeq{\end{equation}}
\def\bea{\begin{eqnarray}}
\def\eea{\end{eqnarray}}

\def\roughly#1{\mathrel{\raise.3ex\hbox
{$#1$\kern-.75em\lower1ex\hbox{$\sim$}}}}
\def\ls{\roughly<}

\def\ra{\rightarrow}
\def\journal#1#2#3#4{{\it #1} {\bf #2} (#3) #4}
\def\epj{Eur. Phys. J.}
\def\prl{Phys. Rev. Lett.}
\def\pl{Phys. Lett.}
\def\np{Nucl. Phys.}
\def\npps{Nucl. Phys. B (Proc. Suppl.)}
\def\ptp{Prog. Theor. Phys.}
\def\ptps{Prog. Theor. Phys. Suppl.}

\def\pr{Phys. Rev.}

\def\ijmp{Int. Jour. Mod. Phys.}
\def\jp{J. Phys.}
\def\aj{Astrohys. J.}
\begin{document}

\begin{flushright}
%DESY 01-051\\
hep-ph/0104123\\
%April 2001\\
\end{flushright}

\begin{center}

{\Large \bf \centerline{Majorana Neutrinos and Same-Sign Dilepton
Production}
\centerline{at LHC and in Rare Meson Decays}}

\vspace*{1.5cm}

{\large A.~Ali} \vskip0.2cm Deutsches Elektronen Synchrotron DESY,
Hamburg \\ \vspace*{0.3cm} \centerline{ and} \vspace*{0.3cm}
{\large A. V.~Borisov, N. B.~Zamorin} \vskip0.2cm  Faculty of
Physics, M. V. Lomonosov Moscow State University,
\\ Moscow 119992, Russia
\vskip0.5cm {\Large Abstract\\} \vskip3truemm

\parbox[t]{\textwidth}{We discuss same-sign dilepton production mediated
by Majorana neutrinos in high-energy proton-proton collisions
$pp\ra \ell^+ \ell^{\prime + }X$ for $\ell,~ \ell^\prime = e,~
\mu,~ \tau$ at the LHC energy $\sqrt{s}=14$ TeV, and in the rare
decays of $K$, $D$, $D_s$, and $B$ mesons of the type $M^{+}\ra
M^{\prime -}\ell ^{+}\ell ^{\prime+}$. For the $pp$ reaction,
assuming one heavy Majorana neutrino of mass $m_N$, we present
discovery limits in the $\left( m_{N},\left|U_{\ell
N}U_{\ell^\prime N}\right|\right)$ plane where $U_{\ell N}$ are
the mixing parameters. Taking into account the present limits from
low energy experiments, we show that at LHC for the nominal
luminosity $L=100~{\rm fb}^{-1}$ there is no room for observable
same-sign dilepton signals. However, increasing the integrated luminosity by
a factor 30, one will have sensitivity to
heavy Majorana neutrinos up to a mass $m_{N}\ls 1.5~{\rm TeV}$
only in the dilepton channels $\mu\mu$ and $\mu\tau$,
but other dilepton states will
not be detectable due to the already existing strong constraints.
We work out a large number of rare meson decays, both for the
light and heavy Majorana neutrino scenarios, and argue that the
present experimental bounds on the branching ratios are too weak
to set reasonable limits on the effective Majorana masses.}

\end{center}

\thispagestyle{empty}
\newpage
\setcounter{page}{1}

\textheight 23.0 true cm

\section{Introduction}
 Recent results from the
KEK to Kamioka long baseline neutrino experiment (K2K) \cite{K2K}
strengthen the neutrino oscillation interpretation of the
atmospheric neutrino anomaly observed earlier by the
Superkamiokande detector \cite{SuperK}. These, as well as the
solar neutrino deficit measurements reported in a number of
experiments
\cite{solar-kamiokande,solar-sage,solar-gallex,solar-gno} yield
valuable information on the neutrino mass differences and mixing
angles \cite{solarneutrino}. The simplest scheme which accounts
for these results is that in which there are just three light
neutrino mass eigenstates with a mass hierarchy analogous to the
quarks and charged leptons, and the observed phenomena of neutrino
oscillations can be accommodated by a mixing matrix in the lepton
sector \cite{MNS} --- analogous to the well-studied quark rotation
matrix \cite{CKM}. If, however, the LSND result \cite{LSND} is
confirmed, it would imply a fourth, sterile, neutrino $\nu_s$\,,
separated in mass from the other neutrinos by typically $0.4 \div
1$ eV. In that case, there might be even more such (sterile)
neutrinos.

While impressive, and providing so far the only evidence of new
physics, the solar and atmospheric neutrino experiments do not
probe the nature of the neutrino masses, i.e., they can not
distinguish between the Dirac and Majorana character of the
neutrinos. The nature of neutrino mass is one of the main unsolved
problems in particle physics and there are practically no
experimental clues on this issue. If neutrino are Dirac particles,
then their masses can be generated just like the quark and charged
lepton masses through weak SU(2)-breaking via the Yukawa
couplings, $m_D=h_\nu v/\sqrt{2}$, where $h_\nu$ is the Yukawa
coupling and $v=\sqrt{2}\langle \phi^0 \rangle =246$ GeV is the
usual Higgs vacuum expectation value. In that case, to get $\leq
1$ eV neutrino masses, one has $h_\nu \leq 10^{-11}$, which raises
the question about the extreme smallness of the neutrino Yukawa
coupling.

If neutrinos are Majorana particles then their mass term
violates lepton number by two units $\Delta L=\pm 2$ \cite{KGP}.
Being a transition between a neutrino and an antineutrino, it can
be viewed equivalently as the annihilation or creation of two
neutrinos. In terms of Feynman diagrams, this involves the
emission (and absorption) of two like-sign $W$-boson pairs
($W^-W^-$ or $W^+ W^+$). If present, it can lead to a large number
of processes violating lepton number by two units, of which
neutrinoless double beta decay ($\beta \beta_{0\nu}$) is a
particular example. The seesaw models \cite{seesaw} provide a
natural framework for generating a small Majorana neutrino mass
which is induced by mixing between an active (light) neutrino and
a very heavy Majorana sterile neutrino of mass $M_N$. The light
state has a naturally small mass $m_\nu \sim m_D^2/M_N \ll m_D$,
where $m_D$ is a quark or charged lepton mass. There is a heavy
Majorana state corresponding to each light (active) neutrino
state. Typical scale for $M_N$ in Grand unified theories (GUT) is
of order the GUT-scale, though in general, there exists a large
number of seesaw models in which both $m_D$ and $M_N$ vary over
many orders of magnitude, with the latter ranging somewhere
between the TeV scale and the GUT-scale \cite{langacker}.

If $M_N$ is of order GUT-scale, then it is obvious that there are
essentially no low energy effects induced by such a heavy Majorana
neutrino state. However, if $M_N$ is allowed to be much lower, or
if the light (active) neutrinos are Majorana particles, then the
induced  effects of such Majorana neutrinos can be searched for in
a number of rare processes.  Among them neutrinoless double beta
decay, like-sign dilepton states produced in rare meson decays and
in hadron-hadron, lepton-hadron, and lepton-lepton collisions, and
$e \to \mu$-conversions have been extensively studied. (See, e.g.,
the papers: $\beta \beta _{0\nu}$
\cite{moscow-heidelberg,faessler,klapdor}, $K^{+}\ra \pi ^{-}\mu
^{+}\mu ^{+}$ \cite{ng,abad,ls,ls1,zuber1,dib}, $pp\ra\ell^{\pm }
\ell^{\pm }X$ \cite{hsiu}, $pp\ra\ell^{\pm } \ell^{\pm }W^{\mp}X$
\cite{almeida}, $e^{\pm}p\ra \mathop {\nu _{e}} \limits^{\left( {
-} \right)} \ell^\pm \ell^{\prime \pm}X$ \cite{buchm1,flanzetal},
the nuclear $\mu^- \to e^+$ \cite{doietal,simkovic} and $\mu^- \to
\mu^+$ \cite{missimer} conversion.)

Of the current experiments which are sensitive to the Majorana
nature of neutrino, the neutrinoless double beta decay, which
yields an upper limit on the $ee$ element of the Majorana mass
matrix, is already quite stringent. The present best limit posted
by the Heidelberg-Moscow experiment \cite{moscow-heidelberg} is:
$\langle m_{ee} \rangle =\vert\sum_{i}\eta_iU_{ei}^2  m_i\vert <
0.26\, (0.34)$ eV at 68\%\,(95\%) C.L., where $\eta_i$ is the
charge conjugation phase factor of the Majorana neutrino mass
eigenstate $\nu_i=\eta_i\nu_i^c$. Despite some dependence of the
actual limit on the nuclear matrix elements, this limit severely
compromises the sensitivity of future $e^- e^-$ colliders
\cite{sub,london}, as well as of searches in the $e^-e^-$ final
states, such as $p p \to e^- e^- X$, induced by a Majorana
neutrino, discussed here. Hence, it is exceedingly important to
push the $\beta \beta_{0\nu}$-frontier; currently there are
several proposals being discussed in literature, which will
increase the sensitivity to $\langle m_{ee} \rangle$ by one to two
orders of magnitude \cite{faessler,klapdor,danilov}, with the
GENIUS proposal reaching $\langle m_{ee} \rangle \sim 10^{-3}$ eV
\cite{klapdor}. Likewise, precision electroweak physics
experiments severely constrain the mixing elements, namely
$\sum_{N}\vert U_{\ell N}\vert^2$, with $\ell=e,~\mu,~\tau$
\cite{buchm2,nar,nar1}.

Taking into account these constraints, we investigate in this
paper the sensitivity to the Majorana neutrino induced effects
involving same-sign dilepton production. We work out the following
two processes in detail:

(i) Dilepton production in the high-energy proton-proton collision

\beq pp\ra \ell^+ \ell ^{\prime + }X \label{col} \eeq
with $ \ell,~ \ell^\prime = e,~ \mu,~ \tau$ at LHC; and

(ii) in rare meson decays of the type

\beq M^+\ra M^{\prime -} \ell^+ \ell^{\prime +} \label{dec} \eeq
for $M=K,~D,~D_s,~B$.

We obtain discovery limits for heavy Majorana neutrinos involved
in the process (\ref{col}) at the LHC energy $\sqrt{s}=14~{\rm
TeV}$. Using the present limits on the branching ratios of rare
decays (\ref{dec}) we set the upper bounds on effective Majorana
masses. From the existing bounds on the elements of the effective
Majorana mass matrix, the indirect constraints on the branching
ratios in question are deduced.

\section{Dilepton production in high-energy $pp$ collisions}

We have calculated the cross section for the process (\ref{col})
at high energies, \beq \sqrt{s}\gg m_{W}, \label{sw} \eeq via an
intermediate heavy Majorana neutrino $N$ in the leading effective
vector-boson approximation \cite{evb} neglecting transverse
polarizations of $W$ bosons and quark mixing.  We use the simple
scenario for neutrino mass spectrum
\[
m_{N_{1}}\equiv m_{N}\ll m_{N_{2}}< m_{N_{3}},...,
\]
and single out the contribution of the lightest Majorana neutrino
assuming
\[
\sqrt{s}\ll m_{N_2}.
\]
The cross section for the process in question is then
parameterized by the mass $m_N$ and the corresponding neutrino
mixing parameters $U_{\ell N}$ and $U_{\ell ^{\prime }N}$:

\beq \sigma \left( pp\ra \ell ^{+}\ell ^{\prime +}X\right)
=C\left( 1-\frac{1}{2}\delta _{\ell \ell ^{\prime }}\right) \left|
U_{\ell N}U_{\ell ^{\prime }N}\right| ^{2}F\left( E,m_{N}\right)~,
\label{cs} \eeq with
\[
C=\frac{ G_{F}^{4}m_{W}^{6}}{8\pi ^{5}} =0.80~\mbox{fb}~,
\]
and \bea F\left( E,m_{N}\right) =\left( \frac{m_{N}}{m_{W}}\right)
^{2}\int_{z_{0}}^{1}\frac{dz}{z}\int_{z}^{1}\frac{dy}{y}\int_{y}^{1}\frac{dx%
}{x}p\left( x,xs\right) p\left( \frac{y}{x},\frac{y}{x}s\right)
h\left( \frac{z}{y}\right) w\left( \frac{s}{m_{N}^{2}}z\right)~.
\label{F} \eea Here, $z_{0}=4m_{W}^{2}/s $, $E=\sqrt{s}$, and
\[
w\left( t\right)= 2+\frac{1}{t+1}-\frac{2\left( 2t+3\right)
}{t\left( t+2\right) }\ln \left( t+1\right)
\]
is the normalized cross section for the subprocess $W^{+}W^{+}\ra
\ell^{+} \ell ^{\prime + }$~ (in the limit (\ref{sw}) it is
obtained from the well-known cross section for $e^{-}e^{-}\ra
W^{-}W^{-}$ \cite{sub} using crossing symmetry). The function
$h(r)$ defined as
\[
h\left( r\right) = -\left( 1+r\right) \ln r-2\left( 1-r\right)
\]
is the normalized luminosity (multiplied by $r$) of $W^{+}W^{+}$
pairs in the two-quark system \cite{evb}, and

\begin{eqnarray*}
p\left( x,Q^{2}\right) = x\sum_{i}q_{i}\left( x,Q^{2}\right)
=x\left(u+c+t+\bar{d}+\bar{s}+\bar{b}\right)
\end{eqnarray*}
is the corresponding quark distribution in the proton.

\begin{figure}[htb]
\centering
\begin{minipage}[c]{0.5\textwidth}
\centering
\includegraphics[scale=0.4]{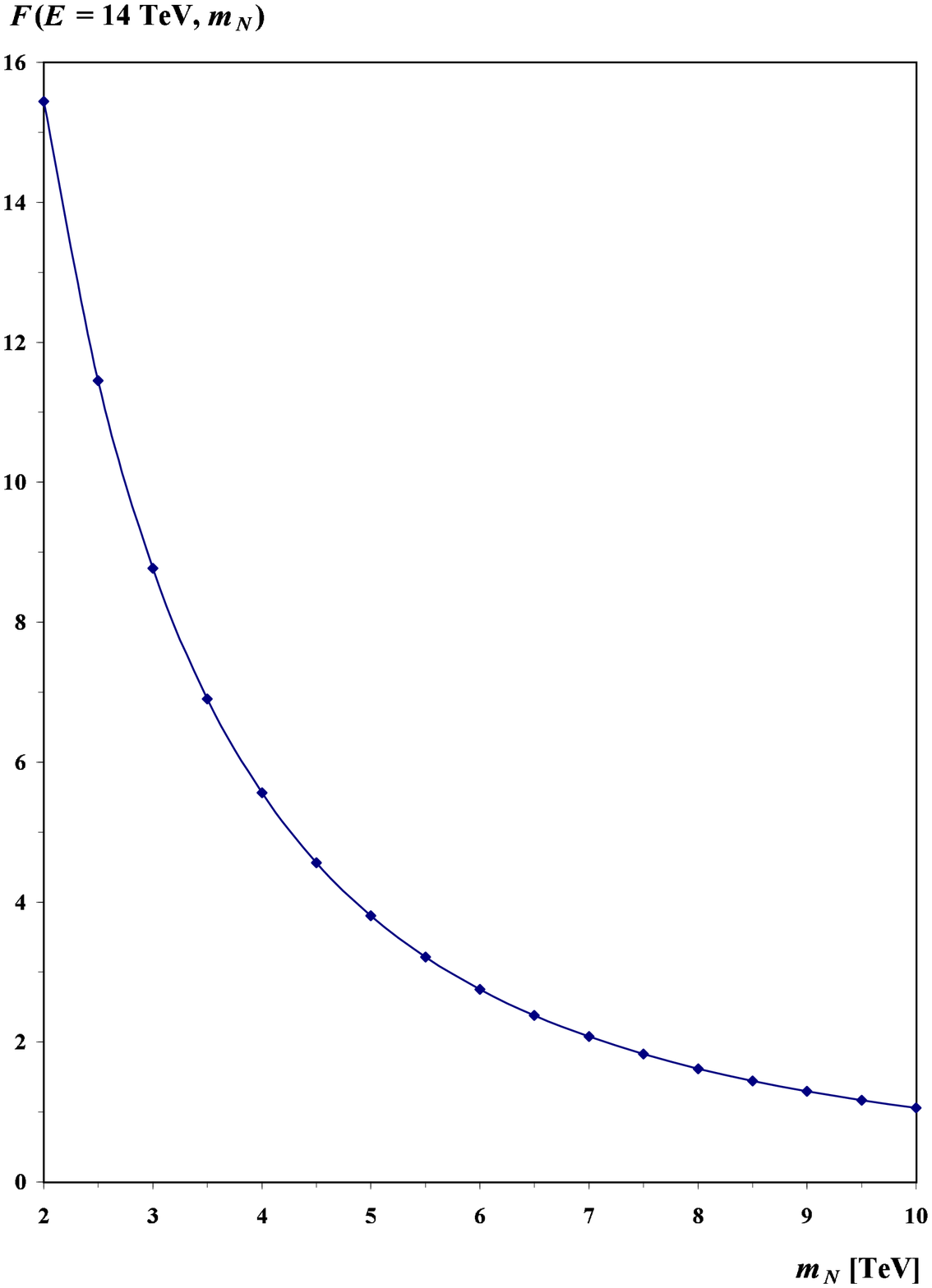}
\end{minipage}
% \hspace*{1cm}
\begin{minipage}[c]{0.49\textwidth}
\centering
\includegraphics[scale=0.42]{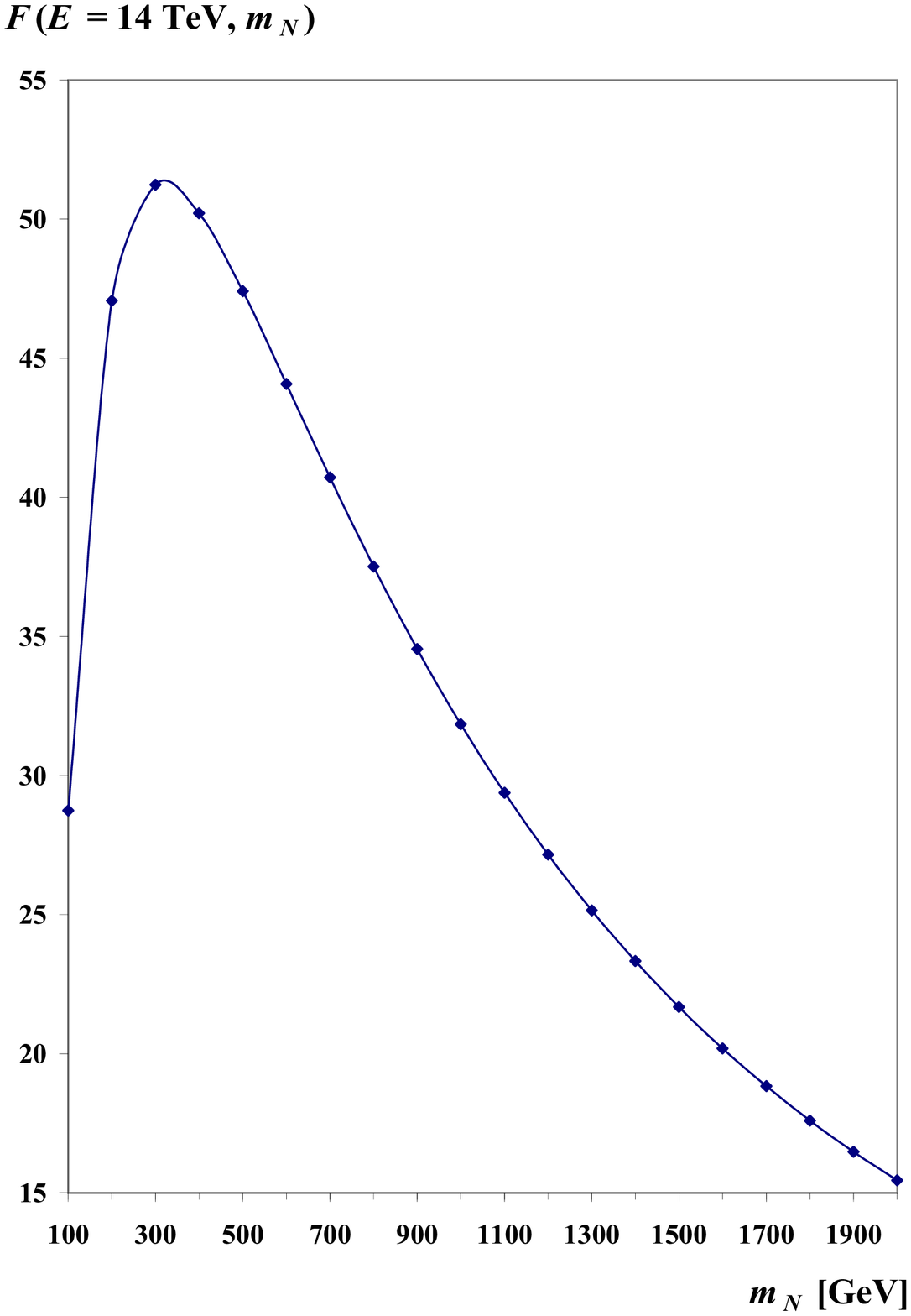}
\end{minipage}
\caption{{\it Left}: The reduced cross section $F(E,m_N)$ defined
in the text for dilepton production as a function of the heavy
Majorana mass $m_N$ at LHC with $E=14$ TeV. {\it Right}: The same
as the left figure but for lighter Majorana neutrinos.}
\label{Fig1}
\end{figure}

In the numerical calculation of the cross section (\ref{cs}) the
CTEQ6 Fortran codes for the parton distributions \cite{cteq} have
been used. The reduced cross section (\ref{F}) as a function of
the neutrino mass $m_N$ is shown in Fig.~\ref{Fig1}  for the LHC
energy $\sqrt{s} = 14~\mbox{TeV}$.

We assume the mixing constraints obtained from the precision
electroweak data \cite{nar} \bea &\sum \left|U_{eN}\right|
^{2}<6.6\times 10^{-3},\quad \sum \left| U_{\mu N}\right|
^{2}<6.0\times 10^{-3} \left(1.8\times 10^{-3} \right),\nonumber
\\ &\sum \left| U_{\tau N}\right| ^{2}<1.8\times
10^{-2}\left(9.6\times 10^{-3} \right).
\label{etau}
\eea
The bound on the mixing matrix elements involving fermions depends on
the underlying theoretical scenario. A mixture of known fermions with new
heavy states (here, a Majorana neutrino) can in general induce both
flavour changing (FC) and non-universal flavour diagonal (FD) vertices
among the light states. The FC couplings are severely constrained for most
of the charged fermions by the limits on rare processes \cite{pdg}. In
Refs.~\cite{nar,nar1}, the FD vertices are constrained by the electroweak
precision data, which we shall use here. There are two limits
obtained on the FD vertices, called in \cite{nar,nar1} the single limit
and joint limit, obtained by allowing just one fermion mixing
to be present or allowing simultaneous presence of all
types of fermion mixings, respectively. The resulting constraints are more
stringent in the single limit case (see numbers in parentheses
on the r.h.s. in Eq. (\ref{etau})) than in the joint limit case, where
the constraints are generally relaxed due to possible accidental
cancellations among different mixings. In our analysis, we shall use the
conservative constraints for the joint limit case.

We must also include the constraint from the double beta decay $\beta
\beta _{0\nu}$, mentioned above.
For heavy neutrinos, $m_N\gg 1~{\rm GeV}$, the bound is \cite{sub}
\beq
\left|\sum_{N(heavy)} U_{e N} ^{2}\frac{\eta_N}{m_{N}}\right|%
<5\times 10^{-5}~{\rm TeV}^{-1}.
\label{beta}
\eeq

In calculating the cross sections for the $\ell\tau $ and
$\tau\tau$ processes, we have used the effective value \beq \left|
U_{\tau
N}\right| _{eff}^{2}={\rm B}_{\tau\mu }\left| U_{\tau N}\right| ^{2}%
<3.1\times 10^{-3} \label{eff} \eeq with ${\rm B}_{\tau\mu }={\rm
Br}\left( \tau ^{-}\ra \mu ^{-}\overline{\nu }_{\mu }\nu _{\tau
}\right) =0.1737$ \cite{pdg}, as this $\tau$-decay mode is most
suitable for the like-sign dilepton detection at LHC (see, e.g.,
\cite{flanzetal}).

Combining the constraints of Eqs. (\ref{etau}), (\ref{eff}), and
(\ref{beta}) and taking a nominal luminosity $L=100~{\rm
fb}^{-1}$, we obtain $\sigma L<1$ for all $\ell\ell^\prime$
channels. Therefore at the LHC there is no room for observable
signals for same-sign dilepton processes $pp\to
\ell\ell^{\prime}X~(\ell, \ell^{\prime}=e,\mu,\tau)$ due to the
existing constraints for the mixing elements $\left| U_{\ell
N}\right|^2$ from the precision electroweak data and neutrinoless
double beta decay.

Let us take into account a recent proposal to increase the
instantaneous LHC luminosity $\cal L$ to a value of $10^{35}~{\rm
cm^{-2}}{\rm s^{-1}}$ \cite{upgrade}, i.e., a total luminosity
$L={\cal L}\times 1~{\rm year} \simeq 3200~{\rm fb}^{-1}$.

\begin{figure}[htb]
\centering
\vspace{1cm}
\begin{minipage}[c]{0.5\textwidth}
\centering
\includegraphics[scale=0.4]{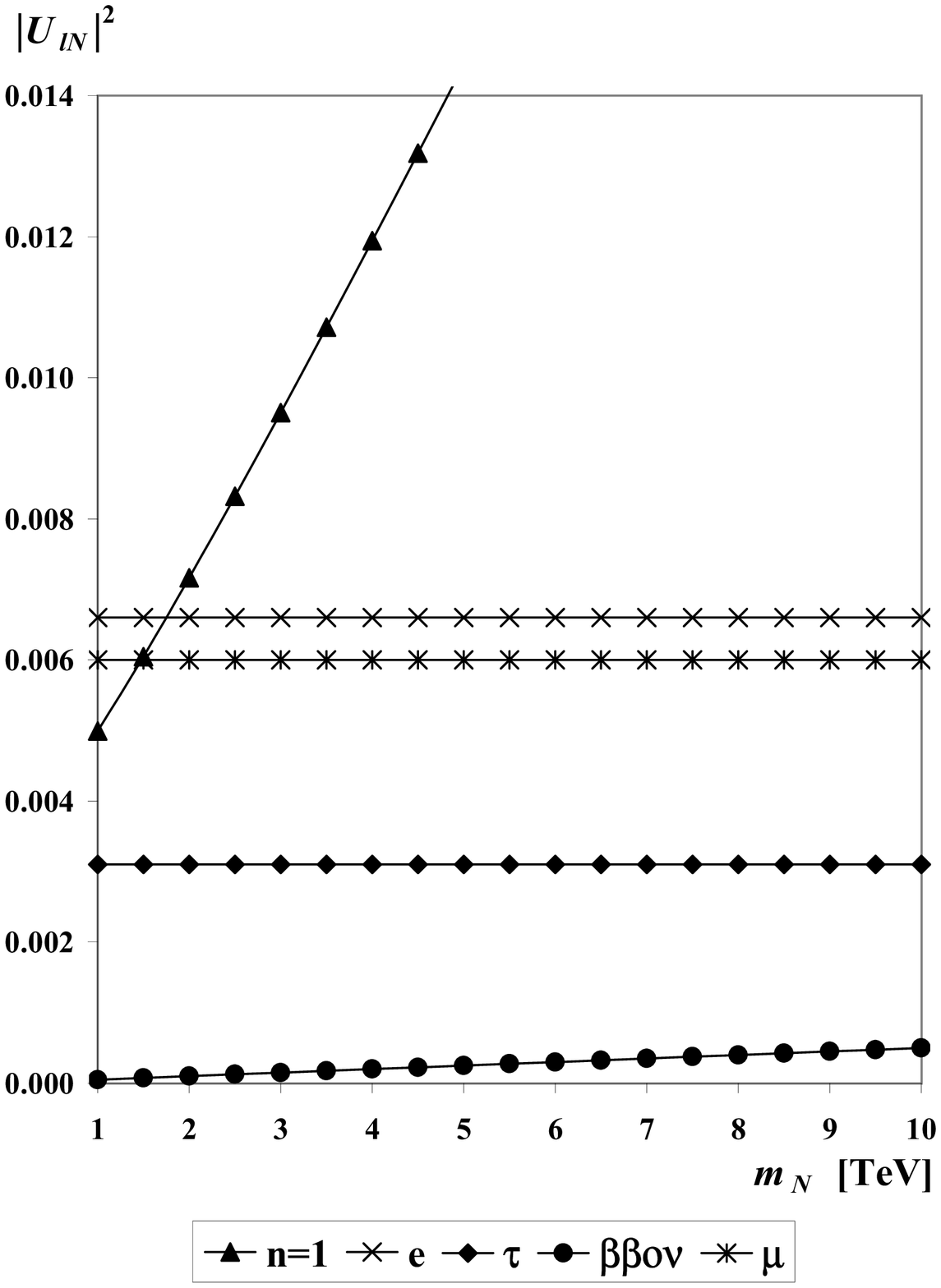}
\end{minipage}
% \hspace*{1cm}
\begin{minipage}[c]{0.49\textwidth}
\centering
\includegraphics[scale=0.4]{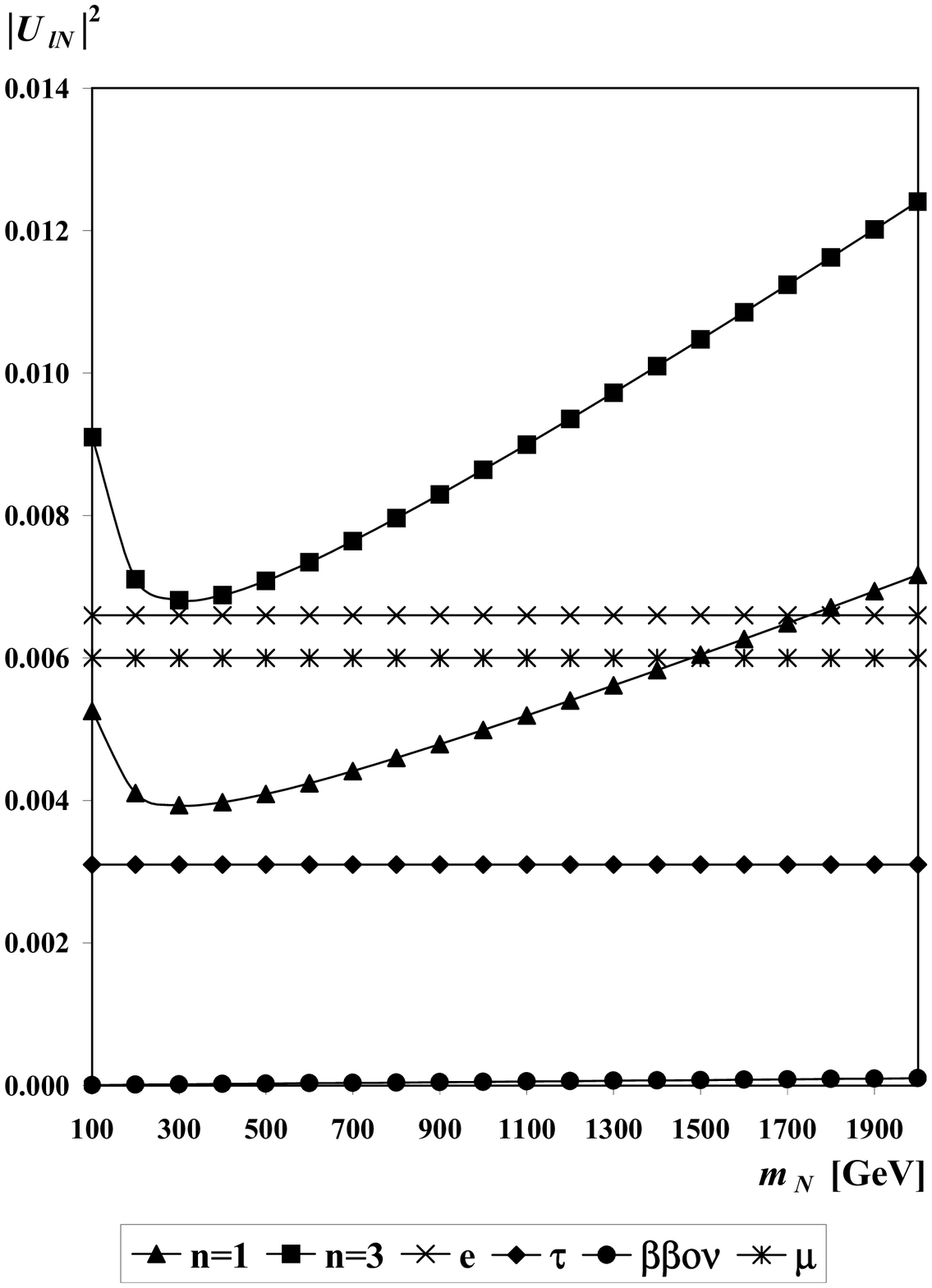}
\end{minipage}
\caption{{\it Left}: Discovery limits for $pp \to \ell^+ \ell^+ X$
as functions of $m_{N}$ and $\left|U_{\ell N}\right|^{2}$ for
$\sqrt{s}=14~\mbox{TeV}$, $L=3200~{\rm fb}^{-1}$ and various values of
$n$, the number of events. We also superimpose the experimental
limit from $\beta \beta _{0\nu}$-decay (Eq.~(\ref{beta})), as well as the
experimental limits on $\left|U_{\ell N}\right|^{2}$ [horizontal
lines for $\ell = e,\,\mu $ (Eq.~(\ref{etau})), and $\tau$
(Eq.~(\ref{eff}))]. {\it Right}:~The same as the left figure~but for
lighter Majorana neutrinos.}
\label{Fig2}
\end{figure}

Using the upgraded LHC luminosity and demanding $n = 1,~3$ events
for discovery (i.e.,~ $\sigma L>n$), we present the
two-dimensional plot for the discovery limits in Fig.~\ref{Fig2}
for the case of identical same-sign leptons ($\ell=\ell^\prime$).

\begin{figure}[htb]
\centering
\begin{minipage}[c]{0.5\textwidth}
\centering
\includegraphics[scale=0.4]{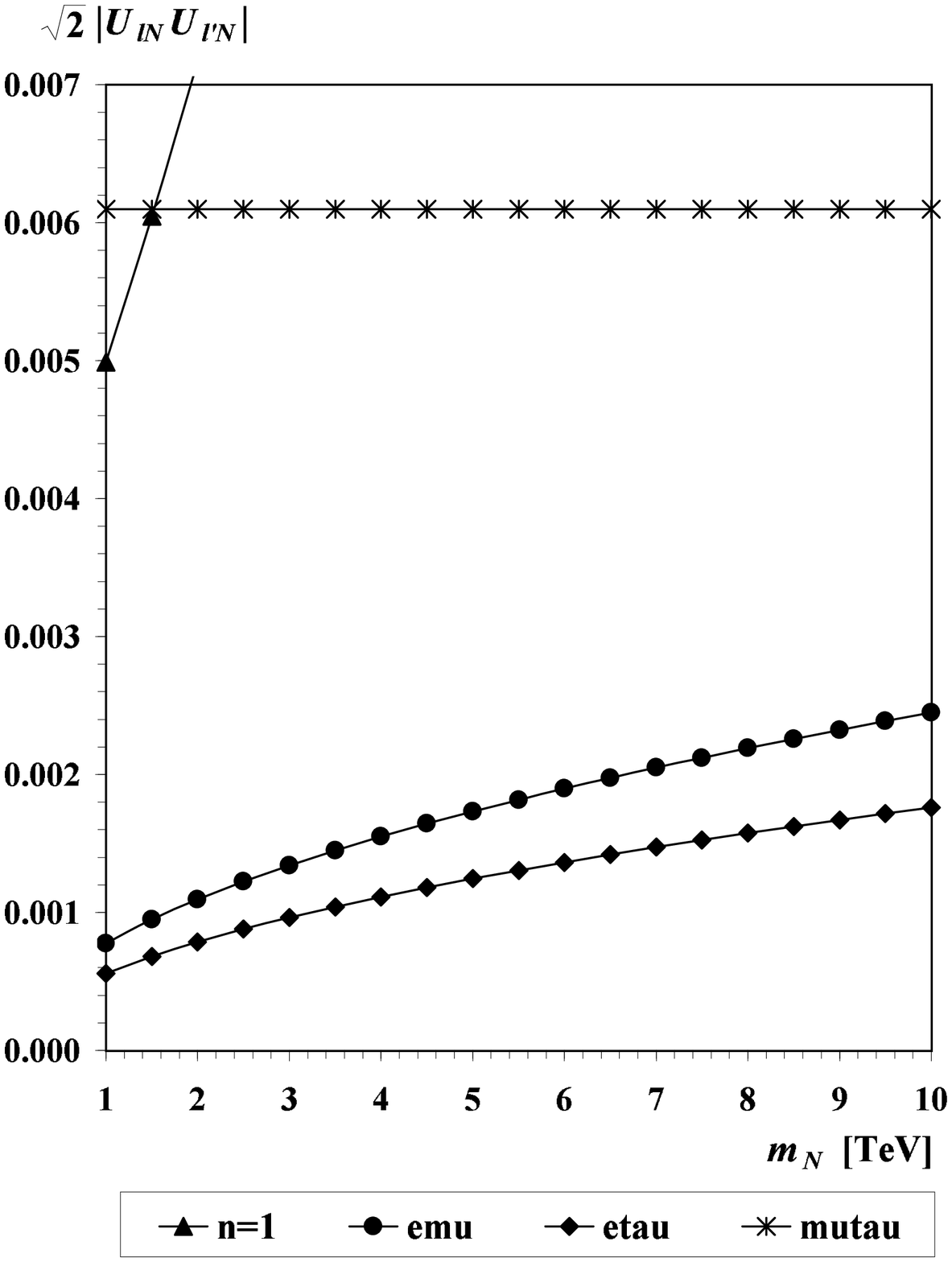}
\end{minipage}
% \hspace*{1cm}
\begin{minipage}[c]{0.49\textwidth}
\centering
\includegraphics[scale=0.4]{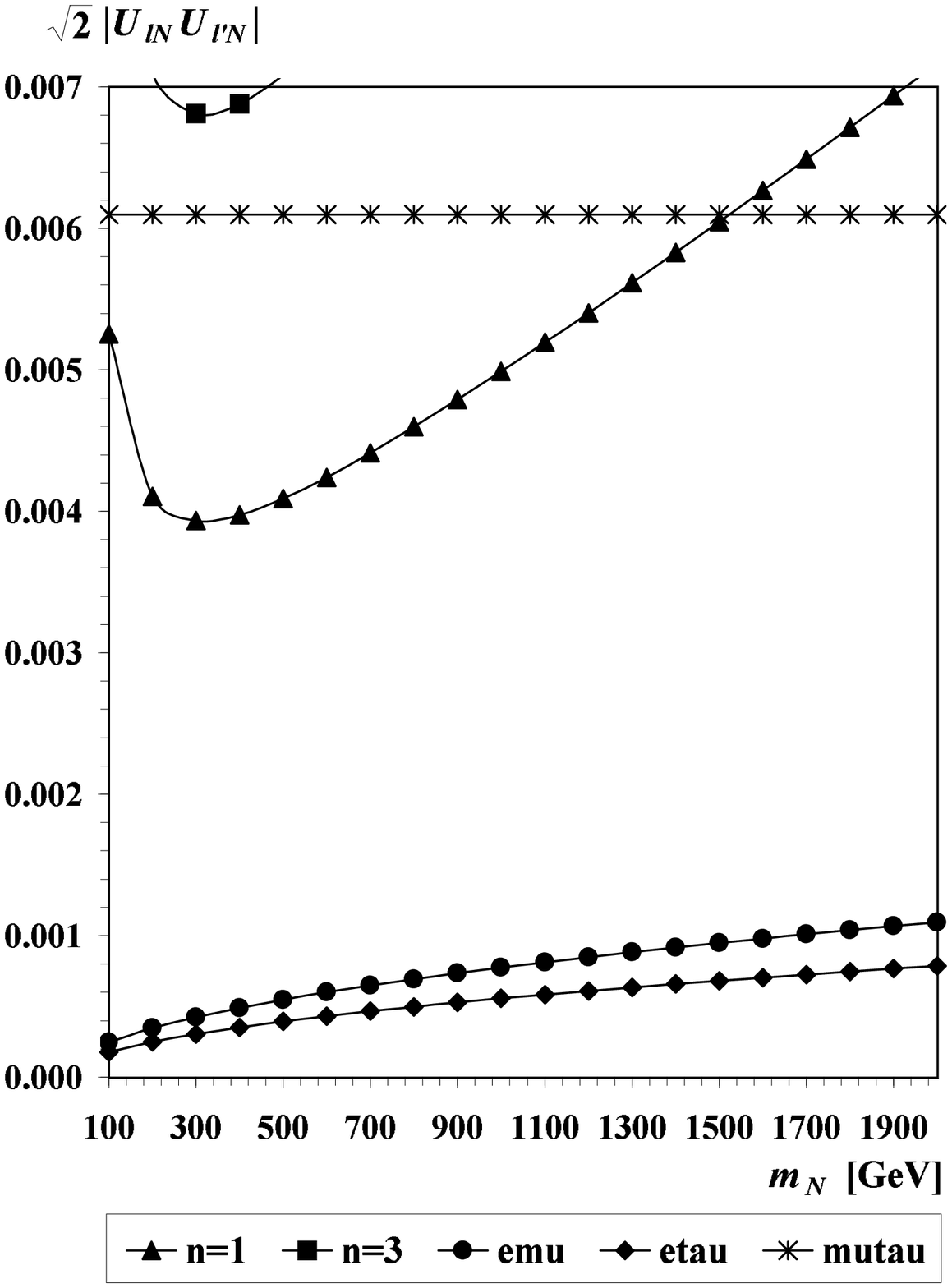}
\end{minipage}
\caption{{\it Left}: Discovery limits for $pp \to \ell^+
\ell^{\prime +} X,~\ell \ell ^{\prime }= e\mu ,~e\tau ,~\mu \tau
$ as functions of $m_N$ and $\sqrt{2}\left|U_{\ell
N}U_{\ell^{\prime} N}\right|$  for
$\sqrt{s}=14~\mbox{TeV}$, $L=3200~{\rm fb}^{-1}$ and various values of
$n$, the number of events. 
 We also superimpose the limits on $\sqrt{2}\left|U_{\ell
N}U_{\ell^{\prime} N}\right|$ obtained from the experimental
limits [Eqs.~(\ref{beta}), (\ref{etau}), and~(\ref{eff})].~{\it
Right}: The same as the left figure but for lighter Majorana
neutrinos.}
\label{Fig3}
\end{figure}

Discovery limits for the case of distinct same-sign leptons, $\ell \ell
^{\prime }= e\mu ,~e\tau ,~\mu \tau $, are shown in
Fig.~\ref{Fig3}.

>From Figs.~\ref{Fig2} and \ref{Fig3} we see that the existing
strong constraints for the mixing elements $\left| U_{\ell
N}\right|^2$ allow a possibility to observe only the same-sign
$\mu\mu$ and $\mu\tau$ processes after increasing the nominal LHC luminosity
$L=100~{\rm fb}^{-1}$ by a factor of about 30. For this case, LHC
experiments will have a sensitivity to heavy Majorana neutrinos of
mass $m_N \ls 1.5~{\rm TeV}$.

Before concluding this section, we would like  to add a comment on
Ref. \cite{flanzetal} where {\it ostensibly} new and improved
bounds on the effective Majorana masses
\beq
\left\langle m_{\ell
\ell^\prime }\right\rangle =\left| \sum_{N}U_{\ell
N}U_{\ell^{\prime} N} m_{N}\eta _{N}\right|
\label{m}
\eeq
have been obtained using the HERA data on $ep$~collisions. The authors
in Ref.~\cite{flanzetal} have assumed that the cross sections for
the processes $e^{\pm}p\ra \mathop {\nu _{e}} \limits^{\left( { -}
\right)} \ell^\pm \ell^{\prime \pm}X$ are proportional to
$\left\langle m_{\ell \ell^{\prime} } \right\rangle^2$. This,
however, is true only for light Majorana neutrinos. For the heavy
Majorana neutrino case, the cross sections depend on the factor
$\left\langle m^{-1}_{\ell \ell{^\prime}}\right\rangle^2$, where
\beq \left\langle m^{-1}_{\ell \ell{^\prime} }\right\rangle =
\left| \sum_{N}U_{\ell N}U_{\ell^{\prime} N} \eta
_{N}\frac{1}{m_N}\right|, \label{m-1} \eeq i.e. the effective {\it
inverse} Majorana masses. As a consequence, Eq. (6) of Ref.
\cite{flanzetal} does not give new physical bounds on the light
Majorana neutrino masses (as the resulting cross section is too
small), and for the heavy Majorana neutrino case, their formula is
not applicable (see also comments on Tables \ref{tab1} and
\ref{tab2} below). Hence, contrary to the claims in
Ref.~\cite{flanzetal}, HERA data does not place any new limit on
the Majorana mass matrix.

\section{Rare meson decays $ M^+\ra M^{\prime -} \ell^+ \ell^{\prime +}$}

\begin{figure}[htb]
\includegraphics[scale=0.91]{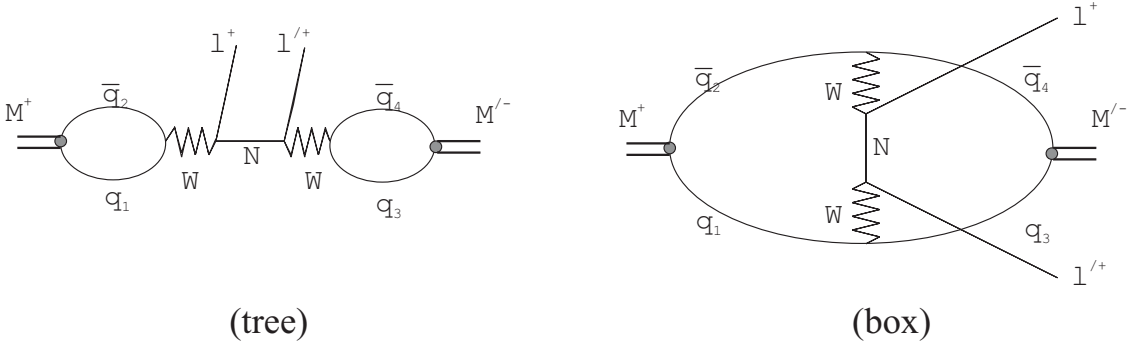}
\caption{Feynman diagrams for the rare meson decay $ M^+\ra
M^{\prime -} \ell^+ \ell^{\prime +}$. Here $N$ is a Majorana
neutrino, bold vertices correspond to Bethe--Salpeter amplitudes
for mesons as bound states of a quark and an antiquark. There are
also two crossed diagrams with interchanged lepton lines.}
\label{Fig4}
\end{figure}

We now take up rare meson decays of mesons of the type (\ref{dec})
mediated by Majorana neutrinos. We shall take the mesons in the
initial and final state to be pseudoscalar. The lowest order
amplitude of the process is given by the sum of the tree  and the
box diagrams shown in Fig.~\ref{Fig4} taken from Ref. \cite{dib}
(for earlier work, see Refs.~\cite{dib,ng,ls}).

It is well-known that the tree diagram amplitude can be expressed
in a model independent way in terms of the measured decay
constants of the pseudoscalar mesons in the initial and final
state, $f_M$ and $f_{M^\prime}$. On the other hand, the box
diagram depends in general on the details of hadron dynamics. In
the usual folklore, the tree diagram dominates and is often used
to set limits on effective Majorana masses \cite{zuber1}.
Calculations based on various quark models for mesons have shown
that this dominance really holds in the decay $K^+\ra\pi^-
\mu^+\mu^+$ for rather small neutrino masses \cite{ng}. It can be
explained, at least partly, by a colour suppression factor
$1/N_c=1/3$ present in the box amplitude. Indeed, we see
in Fig.~\ref{Fig4} that in the quark loops in the tree diagram the
colour summation takes place independently in the two loops. Not so
in the box diagram, where the colour of the quark $q_1$ (antiquark
$\overline{q}_2$) must be the same as the colour of the quark $q_3$
(antiquark $\overline{q}_4$) in the lower (upper) part of the box
diagram. We also note that in some cases there is Cabibbo
suppression of the tree or box amplitude due to smallness of the
corresponding CKM matrix elements.

In this paper, we calculate branching ratios for rare meson decays
(\ref{dec}) for two limiting cases of heavy and light Majorana
neutrinos. In particular, we find that for the case of heavy
neutrinos, $m_N\gg m_M$, the box contribution to the decay
amplitude can also be expressed through $f_M$ and $f_{M^\prime}$
independent of a specific structure of the Bethe--Salpeter vertex
for the meson in question.

The width of the rare decay $M^{+}(P)\ra M ^{\prime
-}(P^\prime)\ell ^{+}(p)\ell ^{\prime +}(p^\prime)$ is given by
\beq
\Gamma _{\ell \ell^\prime}=\left( 1-\frac{1}{2}\delta _{\ell
\ell ^{\prime }} \right)\int \left( 2\pi \right) ^{4}\delta
^{\left( 4\right) }\left( P^{\prime }+p+p^{\prime }-
P\right)\frac{\left| A_{t}+A_{b}\right|
^{2}}{2m_{M}}\frac{d^{3}P^{\prime }d^{3}p\, d^{3}p^{\prime
}}{2^{3}\left( 2\pi \right) ^{9}P^{\prime 0 }p^{0}p^{\prime 0}}~.
\label{wK}
\eeq
Here  $A_t$ ($A_b$) is the tree(box)-diagram
amplitude expressed in the Bethe--Salpeter formalism of Ref.
\cite{est} as
\beq
A_{i}=\frac{1}{\left( 2\pi \right) ^{8}}\int
d^{4}qd^{4}q^{\prime}H_{\mu \nu }^{(i)}L_{i}^{\mu \nu }~,
\label{Ai} \eeq where the lepton tensor is given by \bea
&\ds L_{i}^{\mu \nu }=\frac{g^{4}}{4}\frac{g^{\mu \alpha }}{p_{i}^{2}-m_{W}^{2}%
}\frac{g^{\nu \beta }}{p_{i }^{\prime 2}-m_{W}^{2}}\sum_{N}U_{\ell
N}U_{\ell^\prime N}m_{N}\eta _{N}\nonumber \\ &\ds \times
\left(\overline{v^c}\left(p\right)
\left[ \frac{\gamma _{\alpha }\gamma _{\beta%
}}{\left( p_{i}-p\right)^{2}-m_{N}^{2}}+\frac{\gamma _{\beta
}\gamma _{\alpha }}{\left( p_{i}-p^{\prime }\right)
^{2}-m_{N}^{2}}\right] \frac{1+\gamma ^{5}}{2}v\left(p^\prime%
\right)\right),
\label{Li}
\eea
and $\eta_N$ are the charge conjugation phase factors;
 $i=t,~b$, implying tree and box contributions,
respectively, and
\[
p_t = P,~ p_t^{\prime} = P^\prime;~ p_b = \frac{1}{2}\left(P -
P^\prime\right) +q^\prime -q,~ p_b^{\prime} = \frac{1}{2}\left(P -
P^\prime\right) -q^\prime +q.
\]
The hadron tensors
\bea
&\ds H_{\mu \nu }^{(t)}={\rm Tr}\left\{
\chi ^P\left( q\right) V_{12}\gamma _{\mu }\frac{1+\gamma
^{5}}{2}\right\} {\rm Tr}\left\{ \overline{\chi
}^{P^{\prime}}\left( q^{\prime}\right) V_{43}\gamma _{\nu
}\frac{1+\gamma ^{5}}{2}\right\},\nonumber \\
&\ds H_{\mu \nu}^{(b)}={\rm Tr}\left\{ \chi ^P\left( q\right) V_{13}\gamma _{\mu
}\frac{1+\gamma ^{5}}{2}\overline{\chi }^{P^{\prime }}\left(
q^{\prime}\right) V_{42}\gamma _{\nu }\frac{1+\gamma ^{5}}{2}\right\}
\label{Hi}
\eea
are expressed in terms of the elements of the CKM
matrix, $V_{jk}$\, (the subscripts on $V$ correspond to the quark
line numbering in Fig.~\ref{Fig4}), and the model-dependent
Bethe--Salpeter amplitudes $\chi ^{P}$ for the mesons \cite{est},
\[
\chi ^{P}\left( q\right) =\gamma^5 \left(1  - \delta_M {\not
P}\right)\varphi (P,q)\,\phi_G~,
\]
where $\delta_M = (m_1+m_2)/M^2$, $M$ is the mass of the meson
having a quark $q_1$ and an antiquark $\overline{q}_2$, $m_{1,2}$
are the quark masses, $q = (p_1 - p_2)/2$ is the quark-antiquark
relative 4-momentum, $P = p_1 + p_2 $ is the total 4-momentum of
the meson; the function $\varphi (P,q)$ is model dependent,~ and
$\phi_G$ is the $SU(N_f)\times SU(N_c)$--group factor. The decay
constant $f_M$ of the meson $M$ is expressed through the amplitude
$\chi ^{P}$ as follows:
\[
if_{M}P^{\mu }=\left\langle 0\left |\overline{q }_{2}\left(
0\right) \gamma ^{\mu }\gamma ^{5}q _{1}\left( 0\right)\right
|M\left( P\right) \right\rangle =-i\sqrt{N_c}\int
\frac{d^{4}q}{\left( 2\pi \right) ^{4}}{\rm Tr}\left\{ \gamma
^{\mu }\gamma ^{5}\chi ^{P}\left( q\right) \right\},
\]

\beq
f_M =4\sqrt{N_{c}}~\delta _{M}\int \frac{d^{4}q}{\left(
2\pi \right) ^{4}}\, \varphi \left( P,q\right),
\label{fM}
\eeq
where the sum over colour indices
is implied. The values of the pseudoscalar coupling constants $f_M$
experimentally measured (for $\pi$ and $K$ \cite{pdg}) and the central
values calculated using lattice QCD (for $D$, $D_s$, and $B$ mesons
\cite{aok}) are shown in Table~\ref{tab0}. We have neglected the errors on
these quantities, as we shall see that this will not compromise our
conclusions in any significant way.
\begin{table}
\caption{The pseudoscalar decay constants $f_M$ for the indicated
mesons}
\begin{center}
\begin{tabular}{|c|c|}\hline
Meson& $f_M$ [MeV]\\ \hline $\pi^-$ &130.7 \\ \hline $K^\pm $
&159.8  \\ \hline $D^+$ &228 \\ \hline $D_s^+$ &251\\ \hline $B^+$
&200 \\ \hline
\end{tabular}
\end{center}
\label{tab0}
\end{table}
For all mesons in question, $m_{M}\ll m_{W}$, and we can use the
leading current-current approximation in the lepton tensors
(\ref{Li}).

Below we consider two limiting cases of heavy and light Majorana
neutrinos.

\bigskip

\subsection*{\it Heavy neutrinos: $m_{N}\gg m_{M}$}

\bigskip

For this case, the tree and box lepton tensors are equal to each
other in the leading order of the expansion in $1/m_W^2$:
\bea
&L_{t}^{\mu \nu } =
L_{b}^{\mu \nu }=-16g^{\mu \nu} L(p,p^\prime)~,\nonumber \\ &\ds
L(p,p^\prime)=G_F^2 \sum_{N}U_{\ell N}U_{\ell^{\prime} N} \eta
_{N}\frac{1}{m_N} \left( \overline{v ^c}\left(p\right)
\frac{1+\gamma^{5}}{2}v\left(p^\prime\right)\right).
\label{Lih}
\eea
>From Eqs. (\ref{Ai}), (\ref{Hi}), (\ref{fM}), and (\ref{Lih})
we obtain the total amplitude of the decay
\bea &A=A_t + A_b =
4K_Vf_Mf_{M^\prime}(P\cdot P^\prime)L(p,p^\prime)~, \nonumber\\
&\ds K_V=V_{12}V_{43}+\frac{1}{N_{c}}V_{13}V_{42}~,
\label{Ah}
\eea
which is model independent in this limit.

Using Eqs. (\ref{wK}) and (\ref{Ah}) we calculate the decay width:
\beq \Gamma _{\ell \ell ^{\prime
}}=\frac{G_{F}^{4}m_{M}^{7}}{128\pi ^{3}} f_{M}^{2}f_{M^{^{\prime
}}}^{2}\left| K_{V}\right| ^{2} \left\langle m_{\ell \ell ^{\prime
}}^{-1}\right\rangle ^{2} \Phi _{\ell \ell ^{\prime }}.
\label{wKh} \eeq Here the effective inverse Majorana neutrino mass
$\left\langle m_{\ell \ell ^{\prime }}^{-1}\right\rangle$ is
introduced in the form of Eq. (\ref{m-1}) and $\Phi _{\ell
\ell^\prime }$ is the reduced phase space integral. For identical
leptons \beq \Phi _{\ell \ell }=\int_{4z_{0}}^{z_{1}}dz\left(
z-2z_{0}\right) \left[ \left( 1-\frac{4z_{0}}{z}\right) \left(
z_{1}-z\right) \left( z_{2}-z\right) \right]
^{1/2}\left(1+z_{3}-z\right) ^{2}. \label{Ph=} \eeq For the case
of $\ell$ and $\ell^\prime$ being distinct leptons, assuming
$m_{\ell^\prime}/m_\ell \ll 1$, in the leading approximation, we
have \beq \Phi _{\ell \ell ^{\prime
}}=2\int_{z_{0}}^{z_{1}}\frac{dz}{z} \left( z-z_{0}\right)
^{2}\left[ \left( z_{1}-z\right) \left( z_{2}-z\right) \right]
^{1/2}\left(1+z_{3}-z\right) ^{2}, \label{Ph<<} \eeq where the
variable of integration is $z=\left( P-P^{\prime }\right)
^{2}/m_{M}^{2}$ and the parameters $z_k$ are defined as
\[
z_{0}=\frac{m_{\ell }^{2}}{m_{M}^{2}}\,, \, z_{1}=\left(
1-\sqrt{z_3}\,\right) ^{2},\, z_{2}=\left(
1+\sqrt{z_3}\,\right) ^{2},\,
z_{3}=\frac{m_{M^{\prime}}^{2}}{m_{M}^{2}}\,.
\]

Using Eqs. (\ref{wKh}), (\ref{Ph=}), (\ref{Ph<<}), Table
\ref{tab0} and the experimental values for the total decay widths of
mesons from \cite{pdg}, we have calculated the branching ratios
\beq {\rm B}_{\ell \ell ^{\prime }}(M)=\frac{\Gamma \left( M^{+}\ra
M^{\prime -}\ell ^{+}\ell ^{\prime +}\right) }{\Gamma \left(
M^{+}\ra all\right) }. \label{Br} \eeq Comparing the results with
experimental bounds on ${\rm B}_{\ell \ell ^{\prime }}$ taken from
\cite{app} (for $K$ decays), \cite{E791} (for $D$ and $D_s$
decays), and \cite{pdg} (for $ D^{+}\ra K^{-}\ell ^{+}\ell
^{\prime +}$, $B^{+}\ra \pi^{-}\ell ^{+}\ell ^{\prime +}$, and
$B^{+}\ra K^{-}\ell ^{+}\ell ^{\prime +}$ decays), the upper
bounds for the effective inverse Majorana masses (\ref{m-1}) have
been obtained. The results are shown in Table \ref{tab1}.

\begin{table}
\vspace{-1.5cm} \small{ \caption{Bounds on $\left\langle m_{\ell
\ell ^{\prime }}^{-1}\right\rangle^{-1}$ and indirect bounds on
the branching ratios ${\rm B}_{_{\ell \ell ^{\prime }}}(M)$ for the rare
meson decays $M^{+}\ra M^{\prime -}\ell^{+}\ell ^{\prime +}$ mediated by
Majorana neutrinos (with $m_N \gg m_M)$ and present experimental bounds}
\begin{center}
\begin{tabular}{|c|c|c|c|c|}
\hline Rare decay&Exp. upper bounds &Theor. estimate for& Bounds
on&Ind. bounds
\\ &on ${\rm B}_{_{\ell \ell ^{\prime }}}(M)$& ${\rm
B}_{_{\ell \ell^{\prime }}}(M)/\left\langle m_{\ell \ell ^{\prime
}}^{-1}\right\rangle ^{2}~[{\rm  MeV}^{2}]$& $\left\langle m_{\ell
\ell ^{\prime }}^{-1}\right\rangle^{-1}~[{\rm keV}]$& on ${\rm
B}_{_{\ell \ell ^{\prime }}}(M) $\\ \hline

$K^{+}\ra \pi ^{-}e^{+}e^{+}$ & $6.4\times 10^{-10}$ & $8.6\times
10^{-10}$ & $1200$ &$2.2\times 10^{-30}$ \\ \hline

$K^{+}\ra \pi ^{-}\mu ^{+}\mu ^{+}$ & $3.0\times 10^{-9}$ &
$2.5\times 10^{-10}$ & $300$&$3.5\times 10^{-20}$ \\ \hline

$K^{+}\ra \pi ^{-}e^{+}\mu ^{+}$ & $5.0\times 10^{-10}$ &
$8.4\times 10^{-10}$ & $1300$&$1.2\times 10^{-19}$ \\ \hline
%%%%%%%%%%%%%%%%%%%%%%%%%%%%%%%%%%%%%%%%%%%%%%%%%%%%%%%%%%%%%%%%%%
$D^{+}\ra \pi ^{-}e^{+}e^{+}$ & $9.6\times 10^{-5}$ & $2.2\times
10^{-9}$ & $4.8$&$5.5\times 10^{-30}$ \\ \hline

$D^{+}\ra \pi ^{-}\mu ^{+}\mu ^{+}$ & $1.7\times 10^{-5}$ &
$2.0\times 10^{-9}$ & $11$&$2.8\times 10^{-19}$ \\ \hline

$D^{+}\ra \pi ^{-}e^{+}\mu ^{+}$ & $5.0\times 10^{-5}$ &
$4.2\times 10^{-9}$ & $9.2$ &$5.9\times 10^{-19}$ \\ \hline

$D^{+}\ra K ^{-}e^{+}e^{+}$ & $1.2\times 10^{-4}$ & $2.2\times
10^{-9}$ & $4.3$&$5.5\times 10^{-30}$ \\ \hline

$D^{+}\ra K ^{-}\mu ^{+}\mu ^{+}$ & $1.2\times 10^{-4}$ &
$2.1\times 10^{-9}$ & $4.1$&$3.0\times 10^{-19}$ \\ \hline

$D^{+}\ra K ^{-}e^{+}\mu ^{+}$ & $1.3\times 10^{-4}$ & $4.3\times
10^{-9}$ & $5.7$&$6.1\times 10^{-19}$ \\ \hline

$D_{s}^{+}\ra \pi^{-}e^{+}e^{+}$ & $6.9\times 10^{-4}$ &
$1.9\times 10^{-8}$ & $5.2$&$4.8\times 10^{-29}$ \\ \hline

$D_{s}^{+}\ra \pi ^{-}\mu ^{+}\mu ^{+}$ & $8.2\times 10^{-5}$ &
$1.8\times 10^{-8}$ & $15$ &$2.5\times 10^{-18}$\\ \hline

$D_{s}^{+}\ra \pi ^{-}e^{+}\mu ^{+}$ & $7.3\times 10^{-4}$ &
$3.6\times 10^{-8}$ & $7.1$ &$5.1\times 10^{-18}$\\ \hline
%%%%%%%%%%%%%%%%%%%%%%%%%%%%%%%%%%%%%%%%%%%%%%%%%%%%%%%%%%%%%%%%
$D_{s}^{+}\ra K^{-}e^{+}e^{+}$ & $6.3\times 10^{-4}$ & $2.2\times
10^{-9}$ & $1.9$ &$5.5\times 10^{-30}$\\ \hline

$D_{s}^{+}\ra K^{-}\mu ^{+}\mu ^{+}$ & $1.8\times 10^{-4}$ &
$2.0\times 10^{-9}$ & $3.4$&$2.8\times 10^{-19}$ \\ \hline

$D_{s}^{+}\ra K^{-}e^{+}\mu ^{+}$ & $6.8\times 10^{-4}$ &
$4.2\times 10^{-9}$ & $2.5$ &$5.9\times 10^{-19}$\\ \hline
%%%%%%%%%%%%%%%%%%%%%%%%%%%%%%%%%%%%%%%%%%%%%%%%%%%%%%%%%%%%%%%%
$B^{+}\ra \pi ^{-}e^{+}e ^{+}$ & $3.9\times 10^{-3}$ & $(0.3\div
1.9)\times 10^{-9}$ & $0.3\div 0.7$ &$4.8\times 10^{-30}$ \\
\hline

$B^{+}\ra \pi ^{-}\mu^{+}\mu ^{+}$ & $9.1\times 10^{-3}$ &
$(0.3\div 1.9)\times 10^{-9}$ & $0.2\div 0.5$ &$2.7\times
10^{-19}$\\ \hline

$B^{+}\ra \pi ^{-}e^{+}\mu ^{+}$ & $6.4\times 10^{-3}$ & $(0.6\div
3.8)\times 10^{-9}$ & $0.3\div 0.8$ &$5.4\times 10^{-19}$\\ \hline

$B^{+}\ra \pi ^{-}\tau^{+}\tau ^{+}$ & $$ & $(0.2\div 1.2)\times
10^{-9}$ & $$&$1.7\times 10^{-19}$ \\ \hline

$B^{+}\ra \pi ^{-}e^{+}\tau ^{+}$ & $$ & $(1.0\div 6.2)\times
10^{-10}$ & $$ &$4.0\times 10^{-19}$ \\ \hline

$B^{+}\ra \pi ^{-}\mu^{+}\tau ^{+}$ & $$ & $(1.0\div 6.2)\times
10^{-10}$ & $$ &$8.8\times 10^{-20}$\\ \hline

$B^{+}\ra K ^{-}e^{+}e ^{+}$ & $3.9\times 10^{-3}$ & $(0.2\div
1.5)\times 10^{-10}$ & $0.07\div 0.20$ &$3.8\times 10^{-31}$\\
\hline

$B^{+}\ra K ^{-}\mu^{+}\mu ^{+}$ & $9.1\times 10^{-3}$ & $(0.2\div
1.5)\times 10^{-10}$ & $0.05\div 0.13$ &$2.1\times 10^{-20}$\\
\hline

$B^{+}\ra K ^{-}e^{+}\mu ^{+}$ & $6.4\times 10^{-3}$ & $(0.5\div
2.9)\times 10^{-10}$ & $0.09\div 0.21$ &$4.1\times 10^{-20}$\\
\hline

$B^{+}\ra K ^{-}\tau^{+}\tau ^{+}$ & $$ & $(1.3\div 8.4)\times
10^{-12}$ & $$ &$1.2\times 10^{-21}$\\ \hline

$B^{+}\ra K ^{-}e^{+}\tau ^{+}$ & $$ & $(0.7\div 4.4)\times
10^{-11}$ & $$ &$6.2\times 10^{-21}$\\ \hline

$B^{+}\ra K ^{-}\mu^{+}\tau ^{+}$ & $$ & $(0.7\div 4.4)\times
10^{-11}$ & $$ &$6.2\times 10^{-21}$\\ \hline

\end{tabular}
\end{center}
\label{tab1}}
\end{table}

We can also obtain the indirect upper bounds on the branching
ratios using the  constraint (\ref{beta}) from $\beta \beta
_{0\nu}$ on the $\left\langle m^{-1}_{e e}\right\rangle$  element
of the effective inverse mass matrix. For other elements, assuming
one heavy neutrino scenario, we set
\[
\left\langle m^{-1}_{\ell \ell{^\prime}}\right\rangle <(84.1~{\rm
GeV})^{-1},
\]
using the current mass limits on neutral heavy leptons of the
Majorana type \cite{L3}. The corresponding indirect bounds are
shown in the last column of Table \ref{tab1}.

Our estimate for the $K$ decay branching ratio,
\beq {\rm B}_{\mu\mu}(K)\equiv {\rm
B}\left(K^{+}\ra \pi ^{-}\mu ^{+}\mu ^{+}\right) = 2.5\times
10^{-10}~
{\rm MeV}^2\cdot\left\langle m_{\mu \mu}^{-1}\right%
\rangle ^2 \label{Kh} \eeq should be compared with the
corresponding updated result of Dib et al.~\cite{dib}. Transcribed to
our notations, the relevant decay width is given by the following
expression
\[
\Gamma ^{(DGKS)}\left( K^{+}\rightarrow \pi ^{-}\mu ^{+}\mu
^{+}\right) =7.0\times 10^{-32}\cdot m_{K}^{3}\left\langle m_{\mu
\mu}^{-1} \right\rangle ^2,
\]
yielding a branching ratio
\[
{\rm B}_{\mu\mu}^{(DGKS)}(K)=1.6\times 10^{-10}~{\rm MeV}^2
\cdot\left\langle m_{\mu \mu}^{-1}\right\rangle ^2.
\]
We note that including the box diagram gives the correction factor
(see Eq. (\ref{Ah})) $(1+1/N_c)^2 = 16/9$, i.e. the numerical coefficient
$1.6$ in the above equation gets replaced by $2.8$, which is close to the
numerical coefficient
$2.5$ in our Eq. (\ref{Kh}).

In addition, a rough estimate of the branching ratio,
\[
{\rm B}_{\mu\mu}(K) \sim
0.2\times 10^{-(13\pm 2)}\left(\left\langle m_{\mu \mu}^{-1}\right%
\rangle \cdot 100~ {\rm MeV}\right)^2,
\]
obtained in \cite{ls} (and confirmed in \cite{ls1}) is also in
agreement with Eq. (\ref{Kh}).

>From Table \ref{tab1} we see that the present experimental bounds
on the branching ratios of the rare meson decays are too weak to
give interesting bounds on the Majorana mass. From them and Eq.
(\ref{wKh}), we see that the direct bounds on the effective masses
$\left\langle m^{-1}_{\ell \ell{^\prime}}\right\rangle^{-1}$ of
{\it heavy} Majorana neutrinos are much {\it smaller} than the
difference of meson masses, $m_M - m_{M^\prime}$. The indirect
bounds estimated on ${\rm B}_{\ell \ell^\prime}$, as discussed
above, and given in the fifth column are so small that they can
not be realistically tested in any current or planned experiments.

\bigskip

\subsection*{\it Light neutrinos: $m_{N}\ll m_{\ell}\,,~m_{\ell^\prime}$}

\bigskip

In this case, assuming the tree diagram dominance, the lepton
tensor of Eq. (\ref{Li}), neglecting $m_N^2$ in the dominators,
can be approximated as
\[
L_{t}^{\mu \nu }=8G_F^2\sum_{N}U_{\mu N}^{2}m_{N}\eta _{N} \left(
\overline{v ^c}\left(p\right)\left[ \frac{\gamma ^{\mu }\gamma
^{\nu}}{\left( p_{K}-p\right)^{2}}+\frac{\gamma ^{\nu }\gamma
^{\mu}}{\left( p_{K}-p^{\prime }\right)
^{2}}\right] \frac{1+\gamma ^{5}}{2}v\left(p^\prime%
\right)\right).
\]
Using Eq. (\ref{wK}) with $A\simeq A_{t}$, we obtain the width of
the decay (\ref{dec}) for the case of light Majorana neutrinos
\beq \Gamma _{\ell \ell ^{\prime
}}=\frac{G_{F}^{4}m_{M}^{3}}{16\pi ^{3}} f_{M}^{2}f_{M^{^{\prime
}}}^{2}\left| V_{12}V_{43}\right|^{2} \left\langle m_{\ell \ell
^{\prime }}\right\rangle ^{2} \phi _{\ell \ell ^{\prime }},
\label{wKl} \eeq where $\left\langle m_{\ell \ell ^{\prime
}}\right\rangle$ is the effective Majorana mass (\ref{m}). Here
the phase space integral $\phi _{\ell \ell ^{\prime }}$ has a
rather complicated expression but in the  realistic limit of
massless leptons, $m_{\ell}/m_M \ra 0$ and $m_{\ell^\prime}/m_M
\ra 0$, it can be approximated as
\begin{eqnarray*}
&\ds \phi _{\ell \ell ^{\prime }}\simeq \left( 1-\frac{1}{2}\delta
_{\ell \ell ^{\prime }}\right) \varphi \left( z_{3}\right);\quad
\varphi \left( z_{3}\right) =\int_{0}^{z_{1}}dzz\left[ \left(
z_{1}-z\right) \left( z_{2}-z\right) \right] ^{1/2}\\ &\ds
=\left(1-z_{3}\right) \left[ 2z_{3}+\frac{1}{6}\left(
1-z_{3}\right) ^{2}\right] +z_{3}\left( 1+z_{3}\right) \ln z_{3},
\end{eqnarray*}
where $z_k$ are the same as in Eq. (\ref{Ph<<}).

Using Eq. (\ref{wKl}), we have calculated the branching ratios
(\ref{Br}) and obtained the direct upper bounds on the elements of
the effective Majorana mass matrix. The indirect limits on the
branching ratios have been also obtained with use of a rather
stringent constraint on the $ee$ element,
\[
\left\langle m_{ee}\right\rangle < 1.0~{\rm eV},
\]
>from the $\beta \beta _{0\nu}$ Heidelberg-Moscow
experiment \cite{moscow-heidelberg} (see also comments in \cite{bil}) and
a weaker constraint on other matrix elements used in \cite{dib}
\[
\left\langle m_{\ell\ell^{\prime}}\right\rangle < 9~{\rm eV}~,
\]
which has been deduced in the three light neutrino scenario
assuming the upper bound of 3 eV on the mass of the known
neutrino \cite{bar} (more stringent but model dependent bounds
on $\left\langle m_{\ell\ell{\prime}}\right\rangle$ have been
obtained in \cite{rod}).

Our results are shown in Table \ref{tab2}. We note that  for the
$D^{+}\ra K ^{-}\ell ^{+}\ell ^{\prime +}$ decays, the tree
diagram is strongly Cabibbo suppressed and the box diagram must be
included even for the case of light neutrinos. We have obtained
rough estimates of these decay widths assuming that the reduced
(i.e., without CKM and colour factors) tree and box amplitudes are
of the same order and replacing in Eq. (\ref{wKl}) the factor
$\left| V_{12}V_{43}\right|^{2}$ by $\left|K_{V}\right| ^{2}$ (see
Eq. (\ref{Ah})). It gives a numerical correction factor of about
57.

We conclude once again that the present experimental bounds on the
branching ratios of the rare meson decays are too weak to set
reasonable limits on the effective masses of light Majorana
neutrinos.

\begin{table}[h!]
%\vspace{-1.5cm}
\vspace{-1.5cm} \small{ \caption{Bounds on $\left\langle m_{\ell
\ell ^{\prime }}\right\rangle$ and indirect bounds on
the branching ratios ${\rm B}_{_{\ell \ell ^{\prime }}}(M)$ for
the rare meson decays $M^{+}\ra M^{\prime -}\ell^{+}\ell ^{\prime
+}$ mediated by Majorana neutrinos (with $m_N \ll m_{\ell},
m_{\ell^\prime})$ and present experimental bounds}
\begin{center}
\begin{tabular}{|c|c|c|c|c|}
\hline Rare decay&Exp. upper bounds &Theor. estimate for& Bounds
on &Ind. bounds \\ & on ${\rm B}_{_{\ell \ell
^{\prime }}}(M)$& ${\rm B}_{_{\ell \ell ^{\prime }}}(M)/\left\langle
m_{\ell \ell ^{\prime }}\right\rangle ^{2}~[{\rm  MeV}^{-2}]$&
$\left\langle m_{\ell \ell ^{\prime }}\right\rangle $~[{\rm
TeV}]&on ${\rm B}_{_{\ell \ell ^{\prime }}}(M)$\\ \hline
%%%%%%%%%%%%%%%%%%%%%%%%%%%%%%%%%%%%%%%%%%%%%%%%%%%%%%%%%%%%%%%%%%%%%
$K^{+}\ra \pi ^{-}e^{+}e^{+}$ & $6.4\times 10^{-10}$ & $5.1\times
10^{-20}$ & $0.11$&$5.1\times 10^{-32}$ \\ \hline

$K^{+}\ra \pi ^{-}\mu ^{+}\mu ^{+}$ & $3.0\times 10^{-9}$ &
$1.4\times 10^{-20}$ & $0.47$&$1.1\times 10^{-30}$ \\ \hline

$K^{+}\ra \pi ^{-}e^{+}\mu ^{+}$ & $5.0\times 10^{-10}$ &
$6.2\times 10^{-20}$ & $0.09$&$5.0\times 10^{-30}$ \\ \hline
%%%%%%%%%%%%%%%%%%%%%%%%%%%%%%%%%%%%%%%%%%%%%%%%%%%%%%%%%%%%%%%%%%
$D^{+}\ra \pi ^{-}e^{+}e^{+}$ & $9.6\times 10^{-5}$ & $1.2\times
10^{-21}$ & $280$&$1.2\times 10^{-33}$ \\ \hline

$D^{+}\ra \pi ^{-}\mu ^{+}\mu ^{+}$ & $1.7\times 10^{-5}$ &
$1.2\times 10^{-21}$ & $120$ & $9.7\times 10^{-32}$ \\ \hline

$D^{+}\ra \pi ^{-}e^{+}\mu ^{+}$ & $5.0\times 10^{-5}$ &
$2.3\times 10^{-21}$ & $150$&$1.9\times 10^{-31}$ \\ \hline
%%%%%%%%%%%%%%%%%%%%%%%%%%%%%%%%%%%%%%%%%%%%%%%%%%%%%%%%%%%%%%%%
$D^{+}\ra K ^{-}e^{+}e^{+}$ & $1.2\times 10^{-4}$ & $2.3\times
10^{-21}$ & $230$&$2.3\times 10^{-33}$ \\ \hline

$D^{+}\ra K ^{-}\mu ^{+}\mu ^{+}$ & $1.2\times 10^{-4}$ &
$2.3\times 10^{-21}$ & $230$&$1.8\times 10^{-31}$ \\ \hline

$D^{+}\ra K ^{-}e^{+}\mu ^{+}$ & $1.3\times 10^{-4}$ & $4.5\times
10^{-21}$ & $170$&$3.7\times 10^{-31}$  \\ \hline
%%%%%%%%%%%%%%%%%%%%%%%%%%%%%%%%%%%%%%%%%%%%%%%%%%%%%%%%%%%%%%%%
$D_{s}^{+}\ra \pi^{-}e^{+}e^{+}$ & $6.9\times 10^{-4}$ &
$1.5\times 10^{-20}$ &$210$&$1.5\times 10^{-32}$ \\ \hline

$D_{s}^{+}\ra \pi ^{-}\mu ^{+}\mu ^{+}$ & $8.2\times 10^{-5}$ &
$1.5\times 10^{-20}$ & $74$&$1.2\times 10^{-30}$ \\ \hline

$D_{s}^{+}\ra \pi ^{-}e^{+}\mu ^{+}$ & $7.3\times 10^{-4}$ &
$3.1\times 10^{-20}$ & $150$&$2.5\times 10^{-30}$ \\ \hline
%%%%%%%%%%%%%%%%%%%%%%%%%%%%%%%%%%%%%%%%%%%%%%%%%%%%%%%%%%%%%%%%
$D_{s}^{+}\ra K^{-}e^{+}e^{+}$ & $6.3\times 10^{-4}$ & $5.6\times
10^{-22}$ & $1100$&$5.6\times 10^{-34}$ \\ \hline

$D_{s}^{+}\ra K^{-}\mu ^{+}\mu ^{+}$ & $1.8\times 10^{-4}$ &
$5.6\times 10^{-22}$ & $570$& $4.5\times 10^{-32}$ \\ \hline

$D_{s}^{+}\ra K^{-}e^{+}\mu ^{+}$ & $6.8\times 10^{-4}$ &
$1.1\times 10^{-21}$ & $780$&$8.9\times 10^{-32}$ \\ \hline
%%%%%%%%%%%%%%%%%%%%%%%%%%%%%%%%%%%%%%%%%%%%%%%%%%%%%%%%%%%%%%%%%%%%
$B^{+}\ra \pi ^{-}e^{+}e ^{+}$ & $3.9\times 10^{-3}$ & $(0.3\div
1.8)\times 10^{-23}$ & $(1.5\div 3.6) \times 10^4 $&$1.8\times
10^{-35}$ \\ \hline

$B^{+}\ra \pi ^{-}\mu^{+}\mu ^{+}$ & $9.1\times 10^{-3}$ &
$(0.3\div 1.8)\times 10^{-23}$ & $(2.2\div 5.5) \times 10^4 $
&$1.5\times 10^{-33}$\\ \hline

$B^{+}\ra \pi ^{-}e^{+}\mu ^{+}$ & $6.4\times 10^{-3}$ & $(0.6\div
3.6)\times 10^{-23}$ & $(1.3\div 3.3) \times 10^4 $&$2.9\times
10^{-33}$\\ \hline
%%%%%%%%%%%%%%%%%%%%%%%%%%%%%%%%%%%%%%%%%%%%%%%%%%%%%%%%%%%%%%%%%%%%%%%%
$B^{+}\ra \pi ^{-}\tau^{+}\tau ^{+}$ & $$ & $(1.5\div 9.6)\times
10^{-25}$ & $$&$7.8\times 10^{-35}$ \\ \hline

$B^{+}\ra \pi ^{-}e^{+}\tau ^{+}$ & $$ & $(0.4\div 2.4)\times
10^{-23}$ & $$ &$1.9\times 10^{-33}$\\ \hline

$B^{+}\ra \pi ^{-}\mu^{+}\tau ^{+}$ & $$ & $(0.4\div 2.4)\times
10^{-23}$ & $$&$1.9\times 10^{-33}$ \\ \hline
%%%%%%%%%%%%%%%%%%%%%%%%%%%%%%%%%%%%%%%%%%%%%%%%%%%%%%%%%%%%%%%%%%%%%%%%%%
$B^{+}\ra K ^{-}e^{+}e ^{+}$ & $3.9\times 10^{-3}$ & $(0.2\div
1.2)\times 10^{-24}$ & $(0.6\div 1.4) \times 10^5$ &$1.2\times
10^{-36}$\\ \hline

$B^{+}\ra K ^{-}\mu^{+}\mu ^{+}$ & $9.1\times 10^{-3}$ & $(0.2\div
1.2)\times 10^{-24}$ & $(0.9\div 2.2) \times 10^5$ &$9.7\times
10^{-35}$\\ \hline

$B^{+}\ra K ^{-}e^{+}\mu ^{+}$ & $6.4\times 10^{-3}$ & $(0.4\div
2.4)\times 10^{-24}$ & $(0.5\div 1.3) \times 10^5$ &$1.9\times
10^{-34}$\\ \hline
%%%%%%%%%%%%%%%%%%%%%%%%%%%%%%%%%%%%%%%%%%%%%%%%%%%%%%%%%%%%%%%%%%%%%%%%%%%%%
$B^{+}\ra K ^{-}\tau^{+}\tau ^{+}$ & $$ & $(1.0\div 6.1)\times
10^{-25}$ & $$&$4.9\times 10^{-35}$ \\ \hline

$B^{+}\ra K ^{-}e^{+}\tau ^{+}$ & $$ & $(0.2\div 1.2)\times
10^{-24}$ & $$ &$9.7\times 10^{-35}$\\ \hline

$B^{+}\ra K ^{-}\mu^{+}\tau ^{+}$ & $$ & $(0.2\div 1.2)\times
10^{-24}$ & $$&$9.7\times 10^{-35}$ \\ \hline

\end{tabular}
\end{center}
\label{tab2}}
\end{table}

As for the heavy Majorana neutrino case, one of our results,
\beq
{\rm B}\left(K^{+}\ra \pi ^{-}\mu ^{+}\mu ^{+}\right) = 1.4\times
10^{-20}~
{\rm MeV}^{-2}\cdot\left\langle m_{\mu \mu}\right%
\rangle ^2,
\label{Kl}
\eeq is close to the corresponding result
(in our notations) of  Ref. \cite{dib},
\[
\Gamma ^{(DGKS)}\left( K^{+}\rightarrow \pi ^{-}\mu ^{+}\mu
^{+}\right) =4.0\times 10^{-31}\cdot m_{K}^{-1}\left\langle m_{\mu
\mu}\right\rangle ^2,
\]
or
\[
{\rm B}_{\mu\mu}^{(DGKS)}(K)=1.5\times 10^{-20}~{\rm MeV}^2
\cdot\left\langle m_{\mu \mu}^{-1}\right\rangle ^2,
\]
and the minimal value of a rough estimate of Refs. \cite{ls,ls1},
\[
{\rm B}_{\mu\mu}(K)\sim
0.2\times 10^{-(13\pm 2)}\left(\left\langle m_{\mu \mu}\right%
\rangle /100~ {\rm MeV}\right)^2,
\]
gives the same order of magnitude of the branching ratio as Eq.
(\ref{Kl}).

>From Eq. (\ref{Kl}) we obtain the constraint (see Table
\ref{tab2}) $\left\langle m_{\mu \mu}\right\rangle < 470~{\rm
GeV}$, which is almost the same as the one obtained in \cite{zuber1}:
$\left\langle m_{\mu \mu}\right\rangle \ls 500~{\rm GeV}$. But we
have to stress, in contrast to the conclusion of Ref.
\cite{zuber1}, that there is no reasonable limit at all that emerges
>from rare decays, as Eq.
(\ref{Kl}) is valid for $ m_{N}\ll m_{\mu}\simeq 100~{\rm MeV}$,
and since $\left|U_{\mu N}\right|< 1$  by unitarity, the obvious
inequality (see Eq. (\ref{m})) $\left\langle m_{\mu \mu
}\right\rangle < \left| \sum_{N}m_{N}\right|$ holds. We note that
Ref. \cite{zuber1} was also criticized on the same grounds in Refs.
\cite{ls1,dib}.

\section{Conclusion}

We have examined two processes mediated by Majorana neutrinos: the
production of like-sign dileptons $\ell^+ \ell^{\prime +}$ in
proton-proton collisions at the LHC energy and in the rare decays
of $K^+$, $D^+ $, $D_s^+$, and $B^+$  mesons of the type $M^+ \ra
M^{\prime -}\ell^+\ell^{\prime +}$. We find that at LHC for the
nominal luminosity $L=100~{\rm fb}^{-1}$ there is no room for
observable same-sign dilepton signals due to the existing
constraints from the precision electroweak data and neutrinoless
double beta decay.  But increasing the nominal LHC luminosity by a
factor of about 30 will allow a possibility to observe the
same-sign $\mu\mu$ and $\mu\tau$ processes mediated by a heavy Majorana neutrino
of mass $m_N \ls 1.5~{\rm TeV}$. Data from HERA do not have an
impact on the Majorana mass matrix  --- despite claims to the
contrary \cite{flanzetal}. However, precision electroweak data may
lead to more severe constraints on the fermion mixing angles than
worked out in \cite{nar,nar1} and used by us.  As for the rare
meson decays, present direct bounds on their branching ratios are
too weak to set reasonable limits on effective Majorana masses,
$\left\langle m_{\ell \ell ^\prime}\right\rangle $ (for light
neutrinos) and $\left\langle m_{\ell\ell^\prime}^{-1}\right\rangle
^{-1}$ (for heavy ones). Therefore, to have an impact on the
Majorana mass matrix, a very substantial improvement of the
experimental reach on the lepton-number violating rare meson
decays is needed. Conversely, if a same sign dilepton signal is
seen in any of the meson decay channels listed in Tables
\ref{tab1} and \ref{tab2} in foreseeable future, it will be due to
new physics other than the one induced by  Majorana neutrinos,
such as R-parity violating supersymmetry \cite{belayevetal}.

\section*{Acknowledgements}

We thank Christoph Greub, David London and Enrico Nardi for
helpful discussions and communication, Dmitri Zhuridov for
correcting a coefficient in Eq. (\ref{cs}) and Dmitri Peregoudov for
help in numerical calculations.  A.V.B. thanks DESY for its
hospitality and partial support.


\begin{thebibliography}{100}
%

\bibitem{K2K}
S.H. Ahn et al. (K2K Collaboration), preprint hep-ex/0103001.

\bibitem{SuperK}
Y. Fukuda et al. (Superkamiokande Collaboration),
\journal{\prl}{81}{1998}{1562}; \journal{\prl}{85}{2000}{3999}.

\bibitem{solar-kamiokande}
Y. Fukuda et al. (Superkamiokande Collaboration), preprint
hep-ex/0103033; Y. Fukuda et al. (Kamiokande Collaboration),
\journal{\prl}{77}{1996}{1683}.

\bibitem{solar-sage}
V. Gavrin (SAGE Collaboration), \journal{\npps}{91}{2001}{36};
B.T. Cleveland et al. (Homestake  Collaboration),
\journal{\aj}{496}{1998}{505}; J.N. Abdurashitov et al. (SAGE
Collaboration), \journal{\pl}{B328}{1994}{234}.

\bibitem{solar-gallex}
 W. Hampel et al. (GALLEX Collaboration), \journal{\pl}{B477}{1999}{127};
\journal{\pl}{B388}{1996}{364}.

\bibitem{solar-gno}
E. Bellotti (GNO Collaboration), \journal{\npps}{91}{2001}{44}.

\bibitem{solarneutrino}
See, for a recent analysis of the solar neutrino data, J.H.
Bahcall, P.I. Krastev, and A.Yu. Smirnov, preprint hep-ph/0103179.

\bibitem{MNS}
Z. Maki, M. Nakagawa, and S. Sakata,
\journal{\ptp}{28}{1962}{870}.

\bibitem{CKM}
N. Cabibbo, \journal{\prl}{10}{1963}{531}; M. Kobayashi and K.
Maskawa, \journal{\ptp}{49}{1973}{652}.

\bibitem{LSND}
C. Athanassopoulos et al. (LSND Collaboration),
\journal{\prl}{81}{1998}{1774}; \journal{\pr}{C58}{1998}{2511}.

\bibitem{KGP}
For a nice introduction to the physics of Majorana neutrinos, see
B. Kayser, F. Gibrat-Debu, and F. Perrier, {\it The Physics of the
Massive Neutrinos} (World Scientific, Singapore, 1989).

\bibitem{seesaw}
M. Gell-Mann, P. Ramond, and R. Slansky, in {\it Supergravity},
eds. D. Freedman and P. van Nieuwenhuizen (North Holland,
Amsterdam, 1979) 315; T. Yanagida, in {\it Proceedings of the
Workshop on Unified Theory and Baryon Number in the Universe},
eds. O. Sawada and A. Sugamoto (KEK, Tsukuba, Japan, 1979); R.N.
Mohapatra and G. Senjanovic, \journal{\prl}{44}{1980}{912}.

\bibitem{langacker}
See, for example, P. Langacker,
\journal{\npps}{100}{2001}{383}.

\bibitem{moscow-heidelberg} L. Baudis et al.,
\journal{\prl}{83}{1999}{411}.

\bibitem{faessler} A. Faessler and F. Simkovic,
\journal{\jp}{G24}{1998}{2139}.

\bibitem{klapdor}
H.V. Klapdor-Kleingrothaus, \journal{\it Springer Tracts in Modern
Physics}{163}{2000}{69}; preprint hep-ph/0102277.

%\bibitem{i1} K. Zuber, \journal{\prp}{305}{1998}{295}.

\bibitem{ng}
J.N. Ng and A.N. Kamal, \journal{\pr}{D18}{1978}{3412}.

\bibitem{abad}
J. Abad, J.G. Esteve and A.F. Pacheco,
\journal{\pr}{D30}{1984}{1488}.

\bibitem{ls}
L.S. Littenberg and R.E.~Shrock, \journal{\prl}{68}{1992}{443}.

\bibitem{ls1}
L.S. Littenberg and R.E.~Shrock, \journal{\pl}{B491}{2000}{285}.

\bibitem{zuber1} K. Zuber, \journal{\pl}{B479}{2000}{33}.

\bibitem{dib} C. Dib, V. Gribanov, S. Kovalenko, and I. Schmidt,
\journal{\pl}{B493}{2000}{82}.

\bibitem{hsiu} H. Tso-hsiu, C. Cheng-rui, and T. Zhi-jian,
\journal{\pr}{D42}{1990}{2265}; A.~Datta, M. Guchait, and D.P. Roy,
\journal{\pr}{D47}{1993}{961}; A. Ferrari et~al., \journal{\pr}{D62}{2000}{013001}.

\bibitem{almeida} F.M.L. Almeida Jr, Y.A. Coutinho, J.A. Martins Sim\~oes,
and M.A.B. do Vale, \journal{\pr}{D62}{2000}{075004}.

\bibitem{buchm1}
W. Buchm\"uller and C. Greub, \journal{\np}{B363}{1991}{345}.

\bibitem{flanzetal} M. Flanz, W. Rodejohann, and K. Zuber,
\journal{\pl}{B473}{2000} {324} (Erratum,
\journal{\pl}{B480}{2000}{418}).

\bibitem{doietal}
M. Doi, T. Kotani, and E. Takasugi, \journal{\ptps}{83}{1985}{1};
T.S. Kosmas, G.L. Leontaris, and J.D. Vergados, \journal{\it
Progr. Part. Nucl. Phys.}{33}{1994}{397}.

\bibitem{simkovic}
F. Simkovic, P. Domi, S.G. Kovalenko, and A. Faessler, preprint
hep-ph/0103029.

\bibitem{missimer}
J.H. Missimer, R.N. Mohapatra, and N.C. Mukhopadhyay,
\journal{\pr}{D50}{1994}{2067}.

\bibitem{danilov}
M. Danilov et al., \journal{\pl}{B480}{2000}{12}.

\bibitem{sub} G. B\'elanger, F. Boudjema, D. London, and H. Nadeau,
\journal{\pr} {D53}{1996}{6292}.

\bibitem{london}
D. London, preprint UdeM-GPP-TH-99-65, hep-ph/9907419.

\bibitem{buchm2}
W. Buchm\"uller, C. Greub, and H.-G. Kohrs,
\journal{\np}{B370}{1992}{3}.

\bibitem{nar} E. Nardi, E. Roulet, and D. Tommasini, \journal{\pl}{B344}
{1995}{225}; E.~Nardi, private communication.

\bibitem{nar1} E. Nardi, E. Roulet, and D. Tommasini, \journal{\pl}{B327}
{1994}{319}.

\bibitem{evb} S. Dawson, \journal{\np}{B249}{1985}{42}; M.
Chanowitz and M.K. Gaillard, \journal{\pl}{B142}{1984}{85}; G.L.
Kane, W.W. Repko, and W.B. Rolnik, \journal{\pl}{B148}{1984}{367};
I. Kuss and H. Spiesberger, \journal{\pr}{D53}{1996}{6078}.


\bibitem{cteq} J. Pumplin, D.R. Stump, J. Huston, H.L. Lai, P. Nadolsky, W.K.
Tung, preprint hep-ph/0201195.

%\bibitem{mrs} A.D. Martin, R.G. Roberts, W.J. Stirling, and R.S.
%Thorne, \journal{\epj}{C14}{2000}{133}.

\bibitem{pdg} D.E. Groom et al. (PDG Collaboration),
\journal{\epj}{C15}{2000}{1}.

\bibitem{upgrade} F. Gianotti, M.L. Mangano, T. Virdee et al., preprint hep-ph/0204087.

\bibitem{est} G. Esteve, A. Morales, and R. N\'{u}\~{n}es-Lagos,
\journal{\jp}{G9}{1983}{357}.

\bibitem{aok} S. Aoki, \journal{\ijmp}{A15}{2000}{657}.

\bibitem{app} R. Appel  et al., \journal{\prl}{85}{2000}{2877}.

\bibitem{E791} E.M. Aitala  et al. (E791 Collaboration),
\journal{\pl}{B462}{1999}{401}.

\bibitem{L3} M. Acciarri et al. (L3 Collaboration),
\journal{\pl}{B462}{1999}{354}.

\bibitem{bil} S. M. Bilenky, S. Pascoli, and S.T.
Petcov, preprint hep-ph/0102265.

\bibitem{bar} V. Barger, T.J. Weiler, and K. Whisnant,
\journal{\pl}{B442}{1998}{255}.

\bibitem{rod} W. Rodejohann, \journal{\pr}{D62}{2000}{013011}.

\bibitem{belayevetal}
A. Belyaev et al., preprint CERN-TH/2000-213,
FISIST/8-2000/CFIF, hep-ph/0008276.

\end{thebibliography}
\end{document}